\documentclass[doublecol]{epl2}

\usepackage{graphicx} 
\usepackage{color}
\usepackage{amsmath}
\usepackage{amssymb}
\usepackage{wasysym}
\newcommand{\change}[1]{{{{#1}} }}

\begin{document}
\title{Study of three-dimensional crack fronts under plane stress using a phase field model}
\author{Herv\'e Henry}
\institute{Physique de la Mati\`ere Condens\'ee, \'Ecole Polytechnique, CNRS,
91128 Palaiseau Cedex, France}
\abstract{
  The shape of a crack front propagating through a thin sample is studied using
  a phase field model. The model is shown to have a well defined sharp interface
  limit. The crack front is found to be an ellipse with large axis the width of
  the sample and small axis a function of the Poisson ratio and the width of the
  sample. Numerical results also indicate that the front shape is independent of  
  the crack speed and of the sample extension perpendicular to its width.
}
\maketitle
   
   When a crack grows in an elastic medium under increasing load, one would
   expect, according to the Linear Elastic Fracture Mechanics (LEFM) theory,  
   that its speed increases and reaches
   asymptotically the Rayleigh wave speed\cite{Freund}. Experimentally it has
   been found\change{
   that  at small loads (and therefore low crack speeds) the
   theory  accounts well for the crack speed. But  for higher loads (and for
   higher crack speeds) the observed phenomena are not predicted by the LEFM.
   Indeed, above  a critical speed that is approximately equal to half the  Rayleigh
   wave speed, the crack speed as a function of time starts to be
   irregular\cite{Sharon1996,Sharon1995}  and the crack surface presents marks
   indicating the growth of secondary crack branches. The theoretical effort to
   predict this instability in the framework of the LEFM has not been successful
   yet even though it has been possible to  determine a minimal speed below which the
   branching can not occur\cite{marderreview,katzav2007}. While experiments have
   allowed to describe  the departure from the LEFM theory they  have not been
   able to give a clear description of  the mechanisms leading to
   the instabilities. This is mainly  due to the fact that  crack fronts are
   rapidly moving objects over a long distance compared to the scale at which
   the instability is assumed to take place (the process zone). In addition to
   this issue, experiments are  limited to the observation of the surface of the
   sample\cite{Sharon1995,scheibert2010} while it is likely that the instability mechanism occurs in the
   thickness of the sample\cite{Livne2005,Livne2007}. In fact, the theoretical study of
   propagating cracks has been mostly limited to two dimensional cases and there
   is no widely accepted law of motion for the crack front in 3D.}

\begin{figure}
\includegraphics[width=0.48\textwidth]{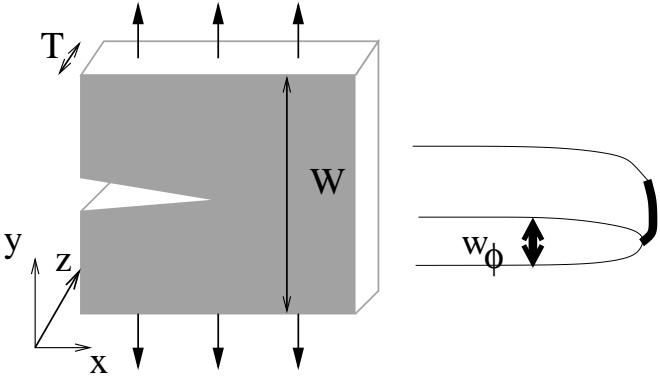}
\caption{\label{figprinciple}\textbf{left:} axis definition and loading conditions. The
boundary conditions are fixed displacement on the top surfaces (xz planes) and
either fixed displacement ($u_z=0$ plane strain)  or zero stress (plane stress) along the z axis on the xy planes. In
the case of stable crack propagation, the crack surface is a part of $y=0$ plane
and the crack front is the boundary of this part in the y plane. \change{\textbf{Left}
Schematic of the isosurface $\phi=0.5$ in the same coordinate system. The crack
front is the thick line that corresponds to the most advanced part of the crack
surface and is a line in the $y=0$ plane. The crack is propagating from left to
right. The isosurface is represented in the material at rest  frame. In order to
observe the usual  parabolic profile  in the $z=0$ plane one simply needs to perform the change of
coordinates: $x\to x+u_x$, $z\to z+u_z$, $y\to y+u_y$ to take into account the
singular strain of the solid at the crack tip.}}
\end{figure}

  \change{ Hence, models  where the breaking mechanism occuring
   in the process zone is described in a simplified way are  appealing since
   their 
   use in numerical simulations would 
   gives access to \textit{real time} observations of the crack front and will
   hopefully allow to have a better understanding of the branching mechanism. 
   One possible candidate would be the  phase field
   model of crack propagation\cite{kkl}. It was originally presented in
   \cite{karma2004} for mode III cracks and then extended to mode I and
   II\cite{Brener2003,henrylevine2004,hakimkarma2005,Pilipenko2007,henry2008}
   and more recently used in the study of the propagation of a tridimensional crack under mixed
   mode loading\cite{Karma2010}.} In this model of
   fracture, elastic energy is converted into surface energy through the
   evolution  of a phase field that governs the elastic constants in the medium
   (for a review of phase field models of crack propagation
   see\cite{spatschek2010}). This approach has 
    the advantages that no law of crack propagation is
   needed (since the crack propagation is due to the evolution of the phase
   field) and that, numerically,  its implementation is straightforward without the need of any complex
   interface tracking approach (this is especially advantageous in 3D where
   surface tracking is involved).\change{ 
   Here,  after briefly presenting
   the model and the numerical setup, I present the study of a  single crack
   front propagating through a thin sample under plane stress conditions
   at its sides(see fig. \ref{figprinciple}). This work aims at solving a long-standing
   question dating back to the  work of Benthem\cite{Benthem1977}
   where it was shown that, contrarily to what is happenning in the quasi
   two-dimensionnal situation of plane strain, a crack front could not be a straight line in a thin
   sample  since the singularity at the crack front combined with the plane
   stress condition would lead to infinite displacement in the direction of the
   front ($z$ direction in fig. \ref{figprinciple}).    Since then, the question of
   the shape of the crack front has remained unanswered due to the particuliar complexity of
   the problem involving \textit{guessing} a law for the crack front motion and
   computing the interaction of the crack front motion and the elastic field.
   Here, in the limit of thin samples, the crack
   front shape (for a given parameter set) is  found to be half an ellipse with large axis the
   thickness of the sample and with small axis a function that is only dependant on
   the thickness of the sample. It should be noted that  it is independant of the crack
   speed and  of the aspect ratio of the sample. When varying the model parameters,
   the value of the ellipse small axis was found not to depend on the dissipation
   at the crack tip but to depend on the Poisson ratio of the elastic material. 
   In addition, in the case of thick samples the crack front was no longer found
   to be an ellipse and its shape was in very good agreement with the prediction
   of Bazant\cite{Bazant79}.  }

   In the phase field model an additional variable, $\phi$ called the phase field
   is introduced. It indicates the internal state of the material. 
   The case $\phi=1$ (resp. $\phi=0$) corresponds to an intact (resp. entirely
   broken) material. The \textit{free energy} from which the evolution equation
   of the phase field and of the elastic field derive  writes:
   \begin{eqnarray}
   \mathcal{F}&=&\iiint dV \frac{w_\phi D}{2}(\nabla
   \phi)^2+\frac{1}{w_\phi}V_{VDW}(\phi)\nonumber\\
   &+&\frac{1}{w_\phi}g(\phi)\left(\frac{\lambda}{2}(\mathrm{tr} \epsilon)^2+\mu
   \mathrm{tr}
   (\epsilon^2)-\epsilon_c^2\right)
   \end{eqnarray}
   where $\epsilon$ is the  symmetric rank 2  strain tensor ($\epsilon_{ij}=(\partial_i
   u_j+\partial_j u_i)/2$) with $u_i$ the displacement field of the material. The
   functions are: $V_{DVW}(\phi)=\phi^2(1-\phi)^2$ and $g(\phi)=4\phi^3-3\phi^4$. 
   $w_\phi$ is a parameter that   sets the interface width (which is proportional
   to $w_\phi$) without changing the fracture energy
   $\gamma=\sqrt{2 w_\phi
   D_\phi}\int_0^1d\phi\sqrt{\frac{1}{w_\phi}(V_{VDW}+(1-g)\epsilon_c^2)}
   $\cite{kkl}. 
   The evolution equation of the displacement  field writes:
   \begin{equation}
    \partial_{tt}u_i=-\frac{\delta \mathcal{F}}{\delta u_i}
   \end{equation}
   so that, if the phase field is uniformly equal to 1, one retrieves the wave
   equation for a solid of density 1 and  that the region where the phase field is
   equal to zero cannot transmit any stress. 
   The evolution equation of the phase field is:
   \begin{equation}
\tau \partial_t \phi=-\frac{\delta \mathcal{F}}{\delta
\phi}/\beta\label{kinetic_phi}
   \end{equation}   
where $\beta$ is a constant kinetic coefficient that measures the dissipation at the
crack tip (see \cite{henry2008}) and  $\tau$ is a variable kinetic coefficient that is equal to 1 except in the
following cases:
\begin{itemize}
\item{it goes to zero if the r.h.s of eq. \ref{kinetic_phi} is
positive, so that the phase field can only decrease (breaking  is 
irreversible).} 
\item{  $\tau$ is  $\max (0, (A+g'(\phi)K_{\mbox{Lam\'e}}(tr \epsilon)^2)/A)$  with
$A=-\delta \mathcal{F}/\delta \phi$ if $tr \epsilon<0$ so that the
compression energy does not contribute to crack growth.}
\end{itemize}
 One should note that
these kinetic modifier, even if they may prevent the system from reaching a
global minimum (in the same way as actual irreversibility does) 
do not lead to an unphysical increase of the free energy. As previously
	mentioned these equations have already been used to describe the growth
	of a crack in a 2D set up. Here I use them to study the crack growth in
	a 3D set-up considering  a plate under plane stress condition
	\begin{equation}
	\sigma_{xz}=\sigma_{yz}=\sigma_{zz}=0 
	\end{equation} 
	which translates in:
	\begin{equation}
	\epsilon_{xz}=0,\,\,\epsilon_{yz}=0\mbox{ and }
	\epsilon_{zz}=-\lambda*(\epsilon_{xx}+\epsilon_{yy})/(\lambda+2\mu)
	\end{equation} 
	with a no-flux boundary condition for the phase field, so that surface
	terms ($\oiint \delta \phi \mathbf{\nabla}\phi.\mathbf{dS}$) do not contribute to the change 
	in the free energy\footnote{Since
	such a boundary condition is not the more generic one for a crack front
	intersecting  the boundary of the material with a finite angle, 
	simulations using  a different boundary condition that allows the
	intersection of the crack front witha an angle were performed without
	noticeable change in the crack front shape.}. 

	The simulations were performed using finite differences (using a scheme
	that derives from a discretized free energy) to compute the
	derivatives and the time stepping was performed using a simple forward Euler
	scheme. The grid spacing was taken equal to $dx=0.3$ and $dx=0.15$ to
	check that  discretisation effects are  negligible. 

	\begin{figure}
\begin{tabular}{cc}
(a)&(b)\\
\includegraphics[width=0.22\textwidth]{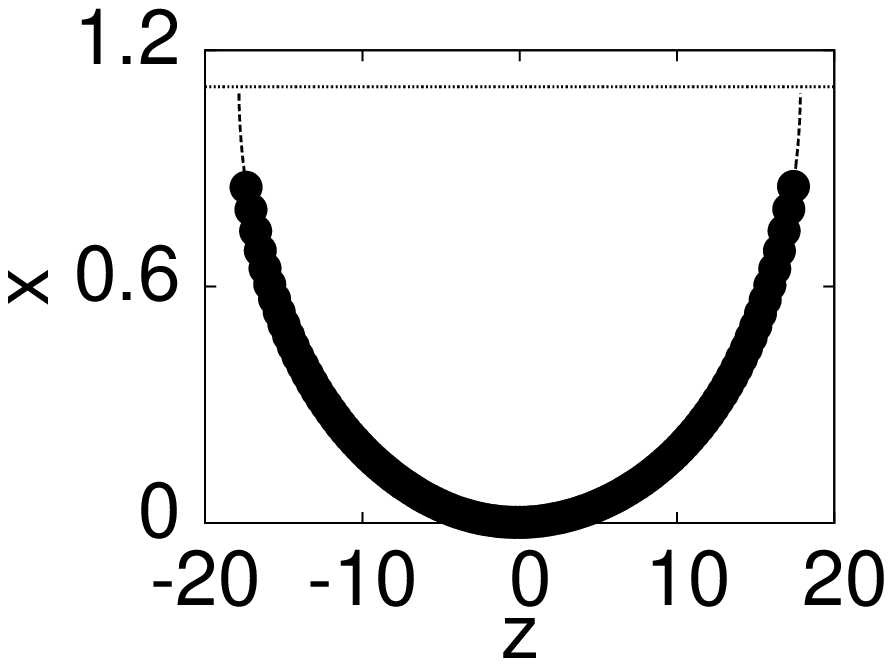}&
\includegraphics[width=0.22\textwidth]{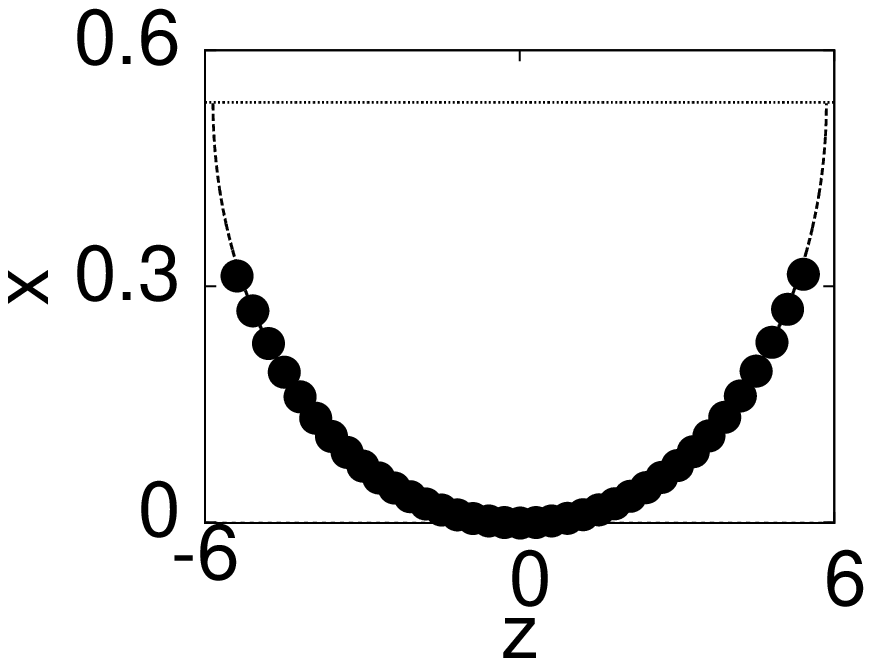}\\
(c)&(d)\\
\includegraphics[width=0.22\textwidth]{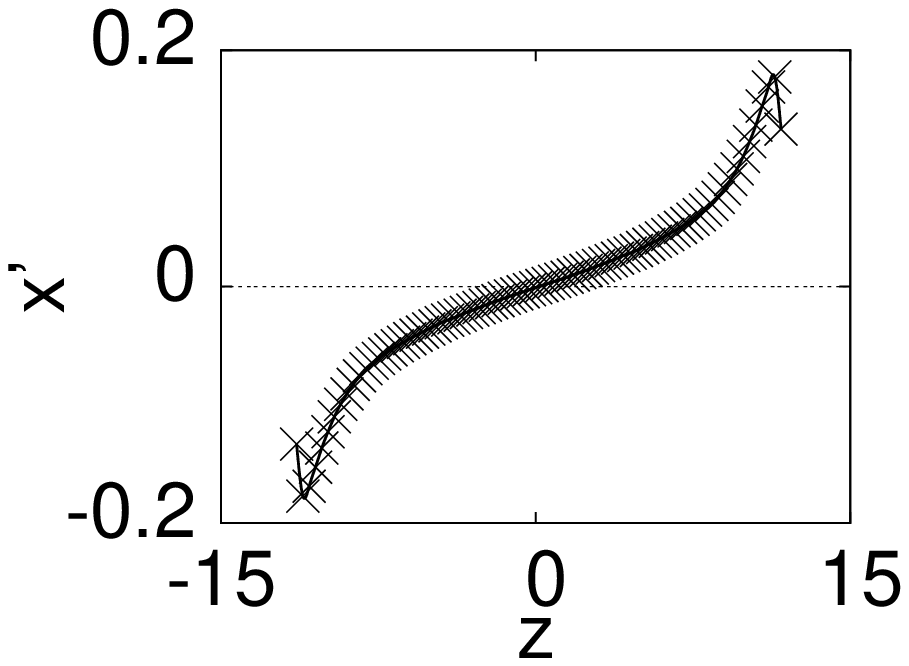}&
\includegraphics[width=0.22\textwidth]{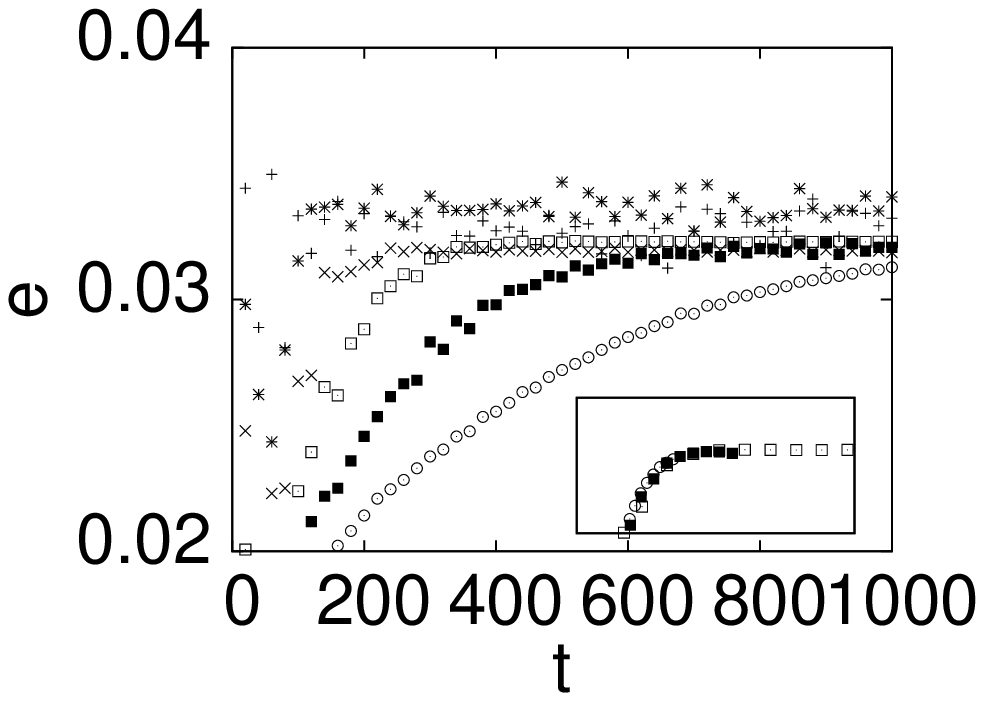}\\
\end{tabular}
\caption{\label{fig_ellips} \textbf{(a)} and \textbf{(b)}$\bullet$ Points of the crack  front computed using the
phase field model. The crack is propagating from top to bottom. The line corresponds to the elliptical fit whose large axis
is the horizontal line. In (a) a small assymetry of the crack front can be
attributed to the use of an asymmetric initial condition. In (b) the crack
front has retrieved a  symmetric shape after a transient regime. 
In both cases, the load was $\Delta_y=22$, leading to a crack speed of 0.51  the
thickness of the
sample was: 36 in \textbf{(a)} and 12 in \textbf{(b)} and the width was 160. Poisson ratio here
was 0.25. The lack of accuracy of the fit close to the boundary can be
attributed to the use of Neuman boundary condition that would impose a crack
front perpendicular to the side of the sample while, an ellipse corresponds to a
tangential crack front. The effect of the boundary condition is clearer when one
considers the slope of the front \textbf{(c)}.\textbf{d} For different values of
$\beta$ (1, 4, 10, 20, 40 and 80), the value of $e$ is plotted as a function of time during the growth of
an accelerating crack. Poisson ratio is 0,25 and the width of the sample is 24.
In the inset, the behavior of $e$ as a function of the distance travelled by
the crack  is shown for $\beta$ equal to 20, 40 and 80.}
\end{figure}

	In this model, the crack surface can be
	seen as the $\phi=0.5$ iso surface and the coordinate system is the
	elastic body at rest. Hence, the crack surface in the $x0y$ plane
	corresponds to a single thin finger using the present coordinate system.
	To retrieve the laboratory frame, one needs to apply the coordinate
	change:$x\leadsto x+u_x(x,y,z),\ y\leadsto y+u_y(x,y,z),\
	z\leadsto z+u_z(x,y,z)$. Using this coordinate change, one retrieves
	the parabolic crack front opening far from the tip (see
	\cite{Bouchbinder2009} for the effects of  a non-linear elastic
	zone at the crack tip). The crack  
	front was taken as the set of  points of higher $x$ of the isoline
	$\phi=0.5$ in each $x0y$ plane (see fig \ref{figprinciple}) and it was a
	line in the $y=0$ plane.
	
	Numerical
	simulations showed that in the case of a single crack the points were
	located at the middle of the sample ($y=0$). While the crack front
	reached a steady shape after a transient regime when the crack speed
	was below the branching threshold, in the quantitative study an
	averaging procedure (usually over 5 to 10  snapshots of the front
	equally spaced in time) was used to reduce the discretisation effects. 

		\begin{figure}
\includegraphics[width=0.48\textwidth]{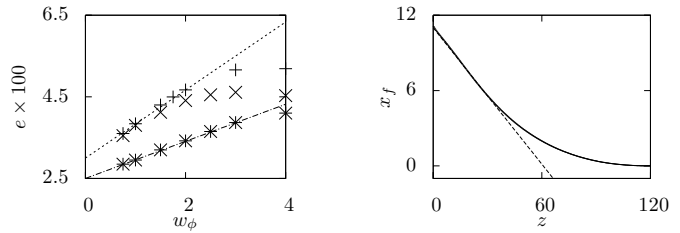}
\caption{\textbf{left}: For a given parameter set and two distinct thickness of the sample $e$ as a function of
$w_\phi$ in space units,
the interface thickness. \label{fig_effetload}. The $+$ and $\times$ correspond to $T=12$ and
$W=80$ and $320$. One can see that in both cases there is convergence toward a
well defined limit that does not depend on $W$. The $*$ symbol corresponds to $T=24$ and
$W=320$. There is also a well defined convergence. The lines serve as  a
guide to the eye. \textbf{right:} shape of half the crack front in the case where
$T/W$ is of order unity (3.4). One can see that it is roughly V-shaped and that
the angle made by the crack front with the free boundary in $z=0$ is roughly
$0.557\pi$ (angle made by  the dashed line), to compare with the value of
$\approx 0.546 \pi$ in fig 10. of
\cite{Bazant79}.}
\end{figure}

        Results are discussed as follows. First, the shape of the crack front in the
       case of thin samples is described. Then the dependence of the crack front on
       the phase field interface thickness is discussed from a quantitative point of
       view and the model is shown to converge toward a \change{well defined
       limit} when the
       interface thickness is decreased toward 0.
      Finally, the dependence of the crack front on relevant physical parameters is
        discussed. \change{The physical parameters considered here are the geometry of
	the sample: that is the ratio $T/W$ and the actual value of $T$, the
	poisson ratio of the material $\nu=\lambda/(2(\lambda+\mu)$, the load applied to the
	material that is directly related to the crack speed  and the kinetic
	coefficient 
	$\beta$ that governs dissipation at the crack tip. }

	 Before describing the crack front, the change in crack speed is
	 briefly discussed. 
	 When considering a single  crack
	 propagating in a thin plate (the quantitative meaning of thin will be precised
	 later), one  expects to retrieve the results one
	 would get from a 2D computation.
	 Indeed,  in both cases, the crack propagation  
	 can be described as transforming elastic energy stored in the
	 material at rest into kinetic energy (in elastic waves) and surface
	 energy. Only a little slow down of the
	 crack due to the additional degree of freedom along the z axis (along
	 which  elastic wave will  also propagate) is expected.
	 Simulations have shown that this is the case and that there is
	 very little difference between the speed computed using 3D simulations
	 and the speed computed using 2D simulations provided the elastic
	 coefficients are rescaled to take into account the zero stress
	 condition. 
	  
	 \begin{figure}
	  \begin{center}
	  \includegraphics[width=0.48\textwidth]{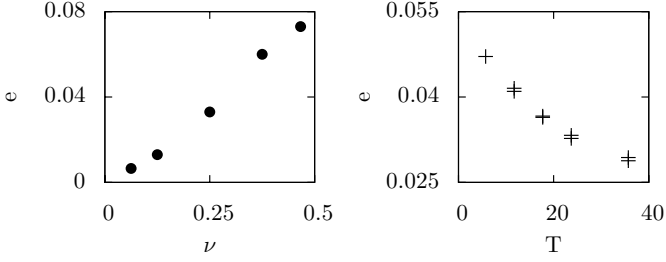}
	  \end{center}
           \caption{\textbf{left:} The ratio $e$ of small axis of the crack front
	   ellipse $eW$ on the sample width $W$ is plotted
           as a function of the Poisson ratio $\nu$. The phase field parameters are
	   kept unchanged. For a given value of Poisson ratio $e$ is taken as
	   the average of a few simulations where the load (and as a result the
	   crack speed) was varied. The variation of $e$ across these
	   simulations, for a given value of $\nu$ was less than 5\% and no
	   clear trend when varying the load could be identified.
	   \textbf{right:} $e$ is plotted as a function of the sample width 
	   for two different values of its extension along y (160 and 320) for a
           value of the poison ratio of 0.25. \label{fig_exentfdenu}
	   }
	  \end{figure}

	 As already shown(\cite{Bazant79}) the zero-stress condition implies that the
	 crack front cannot be flat. Numerical results show that this is
	 actually the case and the crack front is not flat and
	 presents a noticeable curvature at its tip. A fit of the crack front
	 using various possible test functions (such as power laws) indicated  
	 that the best fit is an ellipse with large axis the half thickness of the
	 sample $T/2$ and small axis $eT$ where $e$ is a real constant. (see
	 fig.\ref{fig_ellips}). The equation of the ellipse in the following will
	 write:
         \begin{equation}
	 1=\frac{z^2}{(T/2)^2}+\frac{x^2}{(Te)^2}.\ y=0\label{eqn_ellips}
	 \end{equation}
         It should be noted here that since $T$ is fixed, there is only one
	 adjustable parameter for the fit: $e$ which is \textit{a priori} a function of other
	 parameters describing the system (load, elastic constants and geometry
	 of the sample). 
	 This fit was valid only for thin samples where $W/T$ was approximately larger than 5.
	 For thicker samples the crack front was no longer elliptical and its
	 shape will  be briefly discussed here. \change{I now turn to the
	 convergence of the model, that is the role of the interface thickness
	 $w_\phi$.} 

	 In phase-field simulations, the role of the interface thickness has to
	 be measured and one needs to show that a proper sharp interface limit
	 exists when the interface thickness goes to zero keeping other
	 parameters constant (including the surface energy).  In our model, this
	 can be done by varying $w_\phi$ and one can easily show that the
	 interface thickness is proportional to $w_\phi$ while the surface
	 energy is kept unchanged\cite{kkl}. Simulations using various values of
	 $w_\phi$ (namely 1,2,3 and 4) have shown that the value of $e$
	 converges toward a well-defined limit when $w_\phi$ goes to 0
	 (see  fig. \ref{fig_effetload}).
	 Moreover, they show that the relative error made when considering the case
	 $w_\phi=1$ is of the order  $10\%$.  This indicates that the model has a well
	 defined limit when $w_\phi$ is decreased toward 0. \change{One should
	 note that when $w_\phi$ is decreased both the phase field interface
	 thickness and the phase field \textit{process zone} size are going toward
	 zero and that the model has not a well defined sharp interface limit as
	 solidification models do. Nevertheless, the fact that the size of the
	 \textit{process zone} decreases proportionnally to the interface thickness is
	 coherent with the LEFM theory where the process zone is a point. }
	 In the following all the results
	 considered were  obtained using $w_\phi=1$.
	 
	  \begin{figure}
		 \includegraphics[width=0.4\textwidth]{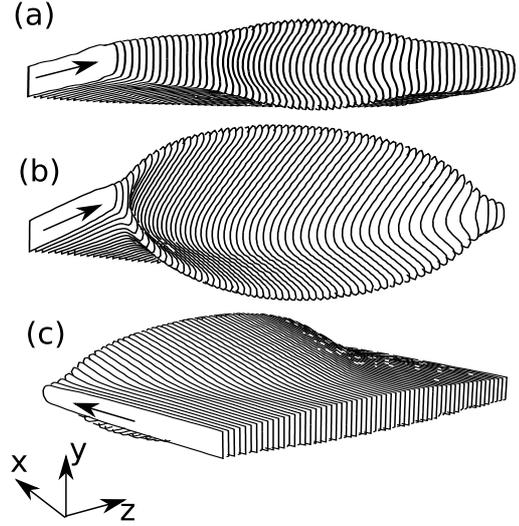}
		 \caption{\label{fig_branching} \textbf{(a)} and
		 \textbf{(b)}\change{ Successive snapshots of the iso
		 surface $\phi=0.5$ taken during a branching event under plane
		 stress condition (more precisely those are juxtaposed isolines
		 taken in each $z=i\ dx$ plane. The whole simulation domain is
		 not shown.). One can see that the branches are appearing
		 at the middle of the sample and spreading toward the
		 sides of the sample. The crack is propagating toward us
		 along the arrow. In \textbf{(c)}, one can see the branching
		 event shown in \textbf{(b)} from behind and the typical shape
		 of the sidebranches that is similar to the branches observed in
		 \cite{Sagi2001}. The schematic axis at bottom correspond to
		 \textbf{(c).}}}
	 \end{figure}

	 I now turn to the study of the dependence of $e$ on the different
	 parameters characterizing the physical system.\change{ First a system where the
	 geometry ($W$ and $T$) , the elastic constants ($\lambda,\ \mu$ and
	 more importantly $\nu$)   and the fracture energy are fixed while
	 the crack speed is varied is considered. The variation in the crack
	 speed can be achieved either by  changing the load or by changing the
	 dissipation ($\beta$) at the crack tip.} As can be seen in
	 fig.\ref{fig_ellips} d, changing the value of $\beta$ does not affect
	 significantly $e$. Varying the load did not change the value of $e$
	 for a wide range of crack speeds (0.05 $c_s$ to 0.4 $c_s$, \change{close to
	 the threshold speed at which the branching instability occurs}).

        In the following, the effects of the parameters 
	describing the mechanical
	 problem (i.e. the Poisson ratio and the geometry of the sample.) are
	 investigated. 

	 First
	 for a given geometry sample (fixed $T$ and $W$), the value of the Poisson ration $\nu$ was
	 varied (by changing $\lambda$ and $\mu$). As shown in  fig. \ref{fig_exentfdenu} the value of $e$ is
	 significantly  affected by
         the Poisson ratio. \change{For small values of $\nu$, the value of $e$
	 is   small and the crack front is  almost flat as it is expected in the
	 plane strain situation. This is not surprising since for $\nu$ close to
	 zero, the amplitude of the  so-called \textit{Poisson effect}  is
	 small  and the difference between plane stress and plane strain
	 condition is small. More precisely, in the limit where  $\nu$ goes to zero, one expects
	 the plane stress and plane strain condition to be equivalent  and therefore the crack
	 front is expected to be flat. Here, the limit of $e$ when $\nu$ goes to
	 zero is zero which is in agreement with this prediction. 

	 On the opposite 
	 for values of $\nu$ close to 0.5 (incompressible limit), the
	 value of $e$ is  much higher and goes toward a finite limit when $\nu$ goes to $0.5$ (its maximal value
	 corresponding to an incompressible material where the difference
	 between the plane strain and plane stress 
	 condition should be the highest). }

         Now I turn to the description of the effects of the sample geometry. All results
	 were obtained for a given Poisson ratio $\nu=0.25$ and, in the case of
	 thin samples,  various values of the width
	 $W$  (80, 160 and 320 space units (su)) and thickness $T$  (from 6 to 60
	 su)  of the sample. The value of the width of the crack $w_\phi$ was
	 kept constant. In the case of thick samples, values of $T$ equal to $W$
	 and larger were used. First the results obtained using thin samples are
	 presented. In this situation, $T\ll W$, the crack front is half an
	 ellipse and it is completely described by the value of $e$. In this
	 situation, it appears that $e$ is a function of $T$ and does not depend
	 on $W$ as can be seen on fig. \ref{fig_exentfdenu} where the computed values of $e$
	 are plotted
	 as a function of  $T$ for two different values of $W$ (160 and 320
	 su) and collapse well on a master curve. When $T$ goes to zero, one can
	 see that $e$ converges toward a well defined finite limit.  When $T$ goes to
	 infinity, the behaviour  of $e$ is still not clear since the present
	 data do not allow to determine wether it converges toward a finite
	 limit or wether it decreases toward 0.
	 
	 When considering  thick samples,  the use of
	 $u_z=u_x=0$ as top and bottom boundary conditions leads to the fact
	 that most of the elastic material is under plane strain.  Then the
	 system is in a situation where  the work of Bazant\cite{Bazant79} can be
	 applied. As a result one expects the crack front to intersect the free
	 boundary with a finite angle that is a function of the Poisson ratio.
	 Numerical results  are in good agreement with this prediction. Indeed,
	 as can be seen in fig. \ref{fig_effetload} where half the crack front
	 is plotted,
	 the crack front is V-shaped and intersects the free boundary with a
	 finite angle. Moreover, the value of the angle obtained during numerical
	 simulations is a function of the
	 Poisson ratio and is in good agreement with the prediction of
	 \cite{Bazant79}.

             \change{   Hence, this work indicates  that the shape of a crack front through a thin
	 sample is half an ellipse that is tangent to its  sides. The small axis
	 of this ellipse is, as one would have expected,  a function of the
	 Poisson ratio of the material and,
	 more surprisingly,   of the thickness of the sample, independantly of its
	 width. Extensive checks on parameters show that the dependance on other
	 parameters of the crack front propagation is not significant. Indeed,
	simulations have shown that the crack front shape is independant of the
	crack speed, the dissipation at the crack tip and the width of the
	sample (provided one stays in the $T \ll W$ regime). In addition, the
	behaviour of the model was checked against predictions made  using the LEFM
	theory in the case of thick samples\cite{Bazant79} and a good agreement
	was found.   

	 This study  stresses one of  the main interest of the phase-field
	modelling of crack propagation in three dimensions: it allows to predict
	the shape of a crack front without any \textit{a priori } hypothesis.

	 Another continuation  of this work would be to use the phase
	field model to understand the branching instability in three dimensions.
	Indeed, from a qualitative point of view, the phase field model allows to
	reproduce well three dimensionnal branching events (see fig.
	\ref{fig_branching}) with a good qualitative agreement with 
	experimental findings\cite{Sagi2001} as far as the shape of the branch
	is concerned.}

  \acknowledgments
I wish to thank Mokhtar Adda-Bedia, Alain Karma, Eran Sharon and Jay Fineberg
for fruitful discussions during this work.


\end{document}